\documentclass[manuscript]{aastex61}

\usepackage{graphicx, amsfonts, amsmath, amssymb,  textcomp, verbatim, listings, subfigure}

\submitjournal{ApJ Supplement Series: Special Issue on Solar/Stellar Astronomy Big Data}

\shorttitle{Period estimation using mutual information for multi-band light curves}
\shortauthors{Huijse et al.}

\watermark{DRAFT}

\begin{document}

\title{Robust period estimation using mutual information for multi-band light curves in the synoptic survey era}

\correspondingauthor{Pablo Huijse}
\email{phuijse@ing.uchile.cl, pablo.huijse@gmail.com}

\author{Pablo Huijse}
\affiliation{Millennium Institute of Astrophysics, Chile} 
\affiliation{Department of Electrical Engineering, Universidad de Chile, Santiago, Chile}
\author{Pablo A. Est\'evez} 
\affiliation{Department of Electrical Engineering, Universidad de Chile, Santiago, Chile}
\affiliation{Millennium Institute of Astrophysics, Chile} 
\author{Francisco F\"orster} 
\affiliation{Center for Mathematical Modeling, Universidad de Chile, Santiago, Chile}
\affiliation{Millennium Institute of Astrophysics, Chile}
\author{Scott F. Daniel}
\affiliation{Department of Astronomy, University of Washington, Seattle, WA, USA}
\author{Andrew J. Connolly}
\affiliation{Department of Astronomy, University of Washington, Seattle, WA, USA}
\author{Pavlos Protopapas}
\affiliation{Institute for Applied Computational Science, Harvard University, Cambridge, MA, USA}
\author{Rodrigo Carrasco}
\affiliation{Department of Electrical Engineering, Universidad de Chile, Santiago, Chile}
\author{Jos\'e C. Pr\'incipe}
\affiliation{Computational Neuroengineering Laboratory of University of Florida, FL, USA}

\begin{abstract}

The Large Synoptic Survey Telescope (LSST) will produce an unprecedented amount of light curves using six optical bands. Robust and efficient methods that can aggregate data from multidimensional sparsely-sampled time series are needed. In this paper we present a new method for light curve period estimation based on the quadratic mutual information (QMI). The proposed method does not assume a particular model for the light curve nor its underlying probability density and it is robust to non-Gaussian noise and outliers. By combining the QMI from several bands the true period can be estimated even when no single-band QMI yields the period. Period recovery performance as a function of average magnitude and sample size is measured using 30,000 synthetic multi-band light curves of RR Lyrae and Cepheid  variables generated by the LSST Operations and Catalog simulators. The results show that aggregating information from several bands is highly beneficial in LSST sparsely-sampled time series, obtaining an absolute increase in period recovery rate up to 50\%. We also show that the QMI is more robust to noise and light curve length (sample size) than the multiband generalizations of the Lomb Scargle and Analysis of Variance periodograms, recovering the true period in 10-30\% more cases than its competitors. A python package containing efficient Cython implementations of the QMI and other methods is provided.

\end{abstract}


\section{Introduction} \label{sec:intro}

The next decade will see the rise of extremely large telescopes \citep{tyson2012}, allowing the astronomers to probe the sky with unprecedented depth, resolution and coverage. An emblematic example of this is the Large Synoptic Survey Telescope (LSST) \citep{ivezic2008lsst, abell2009lsst}. The LSST will begin operations in 2022, capturing the whole southern hemisphere in six bands (ugrizy) over ten years. This translates into 500 PetaBytes of images and 50 PetaBytes in catalogs, corresponding to 37 billion astronomical objects. Robust and computationally-efficient methods are needed in order to process this sheer amount of light curves \citep{tyson2012, feigelson2012big, huijse2014computational}. In this work we focus on the task of period estimation in multi-band light curves such as those that will be produced by the LSST. 
In what follows we describe the problem and review some of the existing methods.

Variable stars are celestial objects whose brightnesses vary through time due to intrinsic or extrinsic reasons \citep{percy2007understanding}. There are certain classes of variable stars, such as Cepheids, RR Lyrae and Eclipsing binaries whose brightnesses vary regularly following periodical patterns. The period of these stars is key in cosmological research as it can be used to measure the distance to their host galaxies. The period of variable stars is also important for asteroseismology research and variable star classification \citep{richards2011machine}. 

The main tool to study variable stars is the \emph{light curve}, a time series of stellar flux or magnitude. Light curves obtained from Earth-based surveys are irregularly sampled due to observation constraints and also have data gaps of different lengths. Light curves are affected by several noise sources, \emph{e.g.} photon noise, sky background noise and scintillation, which can be modeled as uncorrelated (white) noise with variance that changes between samples, \emph{i.e.} light curves have heteroscedastic errors \citep{akritas1997astronomical}. Additionally, light curves are affected by correlated (red) noise due to observations taken with changing air-mass and atmospheric conditions, telescope tracking and other systematics \citep{pont2006effect}. These characteristics make light curve analysis a challenging task. 

Conventional methods for period estimation such as the Fast Fourier Transform (FFT) cannot be directly applied due to the irregular sampling. Several methods have been developed by the statistics and astronomy communities to deal with the analysis of unevenly sampled time series \citep{graham2013comparison}. These methods can be broadly classified as parametric and non-parametric. The most widely used parametric method is the Lomb-Scargle (LS) periodogram \citep{scargle1982}, which equates to finding the best sinusoidal model fit to the light curve in a least squares sense. The LS periodogram has been generalized to take into account heteroscedastic errors \citep{zechmeister2009generalised} and also more complex models based on Truncated Fourier series \citep{palmer2009fast}. 

Phase Dispersion Minimization (PDM) \citep{stellingwerf1978period}, Minimum String Length (MSL) \citep{clarke2002string} and the Analysis of Variance (AoV) periodogram \citep{schwarzenberg1996fast} are classical examples of non-parametric methods. These methods do not rely on sinusoidal models for the data. Instead they optimize a metric on the phase diagram of the light curve $\{\phi_i, m_i\}_{i=1,\ldots,N}$, where $m_i$ are the magnitudes and the phases $\phi_i$ are obtained from the time instants $t_i$ given a certain trial period $P$ as
\begin{equation} \label{eq-folding}
    \phi_i = \frac{\text{mod}(t_i, P)}{P} ~ \in ~ [0, 1],
\end{equation}
where $\text{mod}(\cdot, \cdot)$ stands for the division remainder operator. For example, in the AoV periodogram the phase diagram is binned and a ratio of the variance of the bins and the total variance is computed. By minimizing this ratio over a set of trial periods an estimate of the true period is obtained. Non-parametric methods that rely on information theoretic criteria have also been proposed, {e.g.} The Conditional entropy (CE) periodogram \citep{graham2013using} and the Correntropy Kernelized Periodogram (CKP) \citep{huijse2012information, protopapas2015novel}. In \citep{zucker2016} a statistical criterion for independence using the cumulative distribution of the folded light curve was proposed. This criterion outperformed the Lomb-Scargle in sparsely sampled non-sinusoidal light curves.

LSST \citep{ivezic2008lsst} will produce time series in six optical bands (ugrizy) with non-simultaneous observations \emph{i.e.} time intervals between bands will differ. The main observing strategy consists of two exposures per night for a given field. Fields will be revisited every 3 days on average considering all bands. Single-band average revisit times are longer, \emph{e.g.} r-band is revisited every 15 days. This means that single-band data will be rather sparse. By the end of the first year an average of 18.4 points will be available in the \emph{r}-band. Bands will have different priorities, \emph{e.g.} r-band and i-band will get more visits than the rest. A reliable period detection algorithm for LSST light curves must take into account non-simultaneous observations from all available bands. In recent years \cite{vanderplas2015periodograms} presented an extension of the LS periodogram to sparsely sampled multiband light curves. The multiband LS periodogram combines the single-band periodograms and also fits a term to take into account the variability shared between bands. The AoV periodogram was generalized in a similar way by \cite{mondrik2015multiband}. The multiband AoV is a normalized weighted average of the single-band AoV periodograms. We propose a new period estimation method for multi-band light curves that is based on mutual information. We test this method using synthetic light curves generated using the LSST Operations Simulator (OpSim) and Catalog Simulator (CatSim) \citep{delgado2014lsst, connolly2014, oluseyi2012simulated}. The proposed method achieves better period recovery rates than established methods specially in low sample and low signal-to-noise data. 

\section{Literature review} \label{sec:review}

In this work we make extensive use of the information theoretic concept of Mutual Information (MI). In a broad sense, MI measures the reduction of the uncertainty of a random variable (RV) given that we know a second RV. MI can also be seen as a measure of dependence although, unlike correlation, MI is able to capture non-linear dependence between RVs. More formally, MI is posed as the divergence (statistical distance) between the joint probability density function (PDF) of the RVs and the product of their marginal PDFs. Several definitions of MI exist in the literature, being Shannon's MI the most well known \citep{gray2011entropy}. 
Shannon's MI for continuous RVs $X$ and $Y$ with joint PDF $f_{X, Y}(\cdot, \cdot)$ is defined as 
\begin{align} \label{eq-mis}
   \text{MI}_S(X, Y) &= D_{KL}(f_{X,Y} || f_X f_Y) \nonumber \\
    & = \iint f_{X,Y} \log f_{X,Y} \,dx \,dy - \int f_{X} \log f_X  \,dx  - \int f_{Y} \log f_Y \,dy , 
\end{align}
where $D_{KL}(\cdot || \cdot)$ is the Kullback-Leibler divergence and $f_X (x)= \int f_{X,Y} (x,y)\,dy$, $f_Y (y) = \int f_{X,Y} (x, y)\,dx$ are the marginal PDFs of $X$ and $Y$, respectively. Computing MI using Eq. \eqref{eq-mis} is a difficult task as it requires estimating the joint and marginal PDFs of the RVs. Ideally we want to avoid posing assumptions on the PDFs, hence we focus on non-parametric estimators. Two widely used approaches to compute MI through PDF estimation are the kernel density (KDE) \citep{moon1995} and k-nearest neighbors (KNN) \citep{kraskov2004} estimators. A review of these and other MI estimators using short datasets (50 samples) can be found in \citep{khan2007}.

In this work we intend to avoid the estimation of the PDF by using MI definitions arising from generalized divergences. Such MI estimators have been proposed in the Information Theoretic Learning (ITL) \citep{xu1999, principe2000, principe2010information} literature. In what follows we present the derivation of two MI definitions for continuous RVs from the ITL framework. Starting from the Euclidean distance between probability density functions
\[
D_{ED}(f(x) || g(x)) = \int (f(x) - g(x))^2 \,dx,
\]
the Euclidean distance Quadratic MI \citep{xu1999, principe2010information} between RVs $X$ and $Y$ is defined as 
\begin{align} \label{eq-QMIED}
\text{QMI}_{ED}(X, Y) &= D_{ED}(f_{X,Y} (x,y)|| f_{X}(x) f_{Y}(y)) \nonumber\\ 
&= \iint  f_{X,Y}^2 \,dx \,dy - 2 \iint f_{X,Y} f_X f_Y \,dx \,dy +  \int f_X^2 \,dx \int f_Y ^2 \,dy \nonumber\\
&= V_J - 2 V_C + V_M,
\end{align} 
where $f_{X,Y}(\cdot, \cdot)$ is the joint PDF of $X$ and $Y$ while $f_X(\cdot)$ and $f_Y(\cdot)$ are the marginal PDFs, respectively. 

The terms $V_J$, $V_M$ and $V_C$ correspond to the integrals of the squared joint PDF, squared product of the marginal PDFs and product of joint PDF and marginal PDFs, respectively. The ITL framework provides an estimator of these quantities that can be computed directly from data samples. This estimator is called the Information Potential (IP) \citep{principe2010information} of an RV and it corresponds to the expected value of its PDF.  Note that the expected value of a PDF is equivalent to the integral of the squared PDF. Appendix \ref{sec:ITL} shows how the IP estimator is derived. Assuming that we have $\{x_i, y_i\}_{i=1,\ldots,N}$ i.i.d. realizations of RVs $X$ and $Y$ and using the IP estimator the following expressions are obtained
\begin{equation} \label{eq-vm}
    V_M = \text{IP}_X \text{IP}_Y = \left (\frac{1}{N^2} \sum_{i,j=1}^{N,N}  \text{G}_{\sqrt{2}h} \left( x_i-x_j \right) \right) \left ( \frac{1}{N^2} \sum_{i,j=1}^{N,N} \text{G}_{\sqrt{2}h} \left( y_i-y_j \right) \right),
\end{equation}
\begin{equation} \label{eq-vj}
    V_J = \text{IP}_{X,Y} = \frac{1}{N^2} \sum_{i=1}^{N} \sum_{j=1}^{N} \text{G}_{\sqrt{2}h} \left( x_i-x_j \right)  \text{G}_{\sqrt{2}h} \left( y_i-y_j \right),
\end{equation}
and
\begin{equation} \label{eq-vc}
    V_C = \text{IP}_{X\times Y} = \frac{1}{N} \sum_{i=1}^{N}  \left ( \frac{1}{N}\sum_{j=1}^{N} \text{G}_{\sqrt{2}h} \left( x_i-x_j \right) \right) \left ( \frac{1}{N} \sum_{j=1}^{N} \text{G}_{\sqrt{2}h} \left( y_i-y_j \right) \right),
\end{equation}
where 
\begin{equation}
    \text{G}_{h} \left( x \right) = \frac{1}{\sqrt{2\pi}h} \exp \left( \frac{\|x\|^2}{2h^2} \right),
\end{equation}
is the Gaussian kernel with bandwidth $h$. Note how the integrals have been replaced by sums of pairwise differences between data samples. 

The second ITL quadratic MI that we consider in this work is obtained by defining a divergence measure based on the Cauchy-Schwarz inequality
\[
D_{CS}(f(x) || g(x)) = -\log \frac{\left(\int f(x)g(x) \,dx\right)^2}{\int f(x)^2 \,dx \int g(x)^2 \,dx},
\]
then the Cauchy-Schwarz Quadratic MI \citep{principe2000, principe2010information} for continuous RVs $X$ and $Y$ is
\begin{align} \label{eq-QMICS}
\text{QMI}_{CS}(X, Y) &= D_{CS}(f_{X,Y} (x,y)|| f_{X}(x) f_{Y}(y)) \nonumber\\ 
&= \log \iint f_{X,Y}^2 \,dx \,dy -2 \log \iint  f_{X,Y} f_X f_Y \,dx \,dy  + \log \int f_X^2 \,dx \int f_Y ^2 \,dy \nonumber\\
&= \log V_J - 2 \log V_C + \log V_M,
\end{align} 
where $V_M$, $V_J$ and $V_C$ are computed using Eq. \eqref{eq-vm}, \eqref{eq-vj} and \eqref{eq-vc}, respectively. In the following sections we adapt these QMI estimators for the case of period estimation in light curves.


\section{Methods}

\subsection{Generating synthetic LSST light curves} \label{sec-lsst}

In this section we describe the procedure to generate synthetic light curves using the LSST Operation Simulator (OpSim) and the Catalog Simulator (CatSim) tools. In its normal operation regime the LSST will visit the same field every 3 nights. Six bands will be available (ugrizy). The single-visit $5\sigma$ depth in the r-band will be approximately 24.5. The actual cadence will depend on weather conditions, slew and filter-change times, downtime due to maintenance, among other factors. The OpSim simulates these factors to produce multi-band pointings that are consistent with the LSST scientific drivers. 

The CatSim provides tools to generate different types of sources. In our case we are interested in generating periodic variable stars. CatSim requires the user to specify a normalizing magnitude, spectral energy distribution (SED) and a template. The template file sets the variability type and the period of the resulting light curve. Templates of Cepheids (CEPH), ab-type RR Lyrae (RRab) and c-type RR Lyrae (RRc) are available, among other models. The CatSim RR Lyrae models correspond to \cite{sesar2010light} Stripe 82 Sloan Digital Sky Survey (SDSS) templates. SEDs of modelled main sequence stars are also available. 

The procedure we use to generate a synthetic light curve is
\begin{enumerate}
\item Select a variability model, \emph{e.g.} CEPH, RRab or RRc
\item Randomly select a template associated to the variability type. This defines the period.
\item Randomly select a SED profile.
\item Randomly select a normalizing magnitude by drawing from $U(20, 25)$, where $U(a, b)$ is the Uniform distribution.
\item Randomly select an MJD for the initial phase of the template by drawing from $U(59580, 60580)$.
\item Randomly select a position in the sky by drawing RA from $U(65, 75)$ and DEC from $U(-30, -20)$.
\item Generate the object using CatSim \emph{StellarLightCurveGenerator} class.
\item Generate a set of multi-band pointings for the synthetic object using OpSim according to its position in the sky.
\end{enumerate}
As we are only interested in estimating period recovery, we do not aim to model realistic sky distribution of these variables. A python code that executes this procedure and also the resulting light curves used in this paper can be found at \url{github.com/phuijse/LSST_simulations}. To run this code, previous installation of the LSST simulations framework is required\footnote{Instructions can be found at \url{confluence.lsstcorp.org/display/SIM/Catalog+Simulations+Documentation}}. We run this procedure to generate a set containing 1000 synthetic LSST light curves for each variability type. 

This procedure generates a ``clean'' light curve $\{t_i, m_i, \sigma_i\}_i^b$ with $i=1,\ldots, N_b$, where $t$ corresponds to time instant in MJD, $m$ is the stellar magnitude, $\sigma$ is the photometric error and $b$ denotes the band index. Bands may have a different number of points $N_b$. The photometric error is generated according to Eq. 4 of \citep{ivezic2008lsst}. 

\begin{figure}[t]
\centering
\includegraphics[scale=0.75]{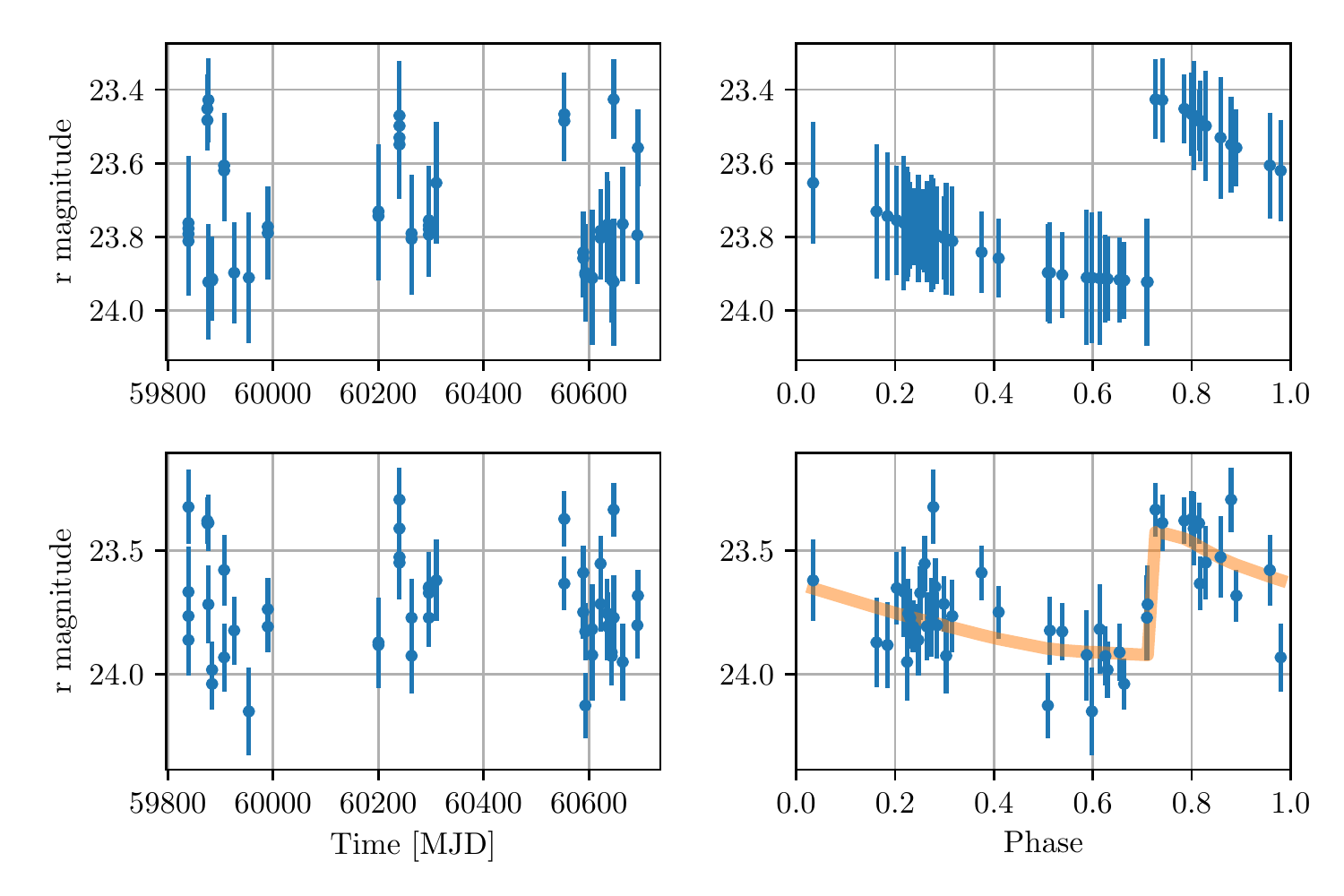}
\caption{\label{fig-examplelc} Synthetic ab-type RR Lyrae light curve with a period of 0.64698 days. The upper row corresponds to the clean light curve obtained using the LSST tools and its respective phase diagram. The lower row corresponds to a contaminated realization of the original light curve. In the lower right plot the line corresponds to the clean light curve.}
\end{figure}

The last step to produce a realistic light curve is to contaminate the clean magnitude values $m$ with the photometric error $\sigma$. This is done by drawing a standard normal RV  $\{r_i\}$ of length $N_b$ and then updating the magnitudes as $\widehat m_i = m_i + r_i \sigma_i$. For each of the ``clean'' light curves we draw 10 contaminated light curves, thus increasing the size of the set to 10000 per variability type. Fig. \ref{fig-examplelc} shows an example of a synthetic RRab light curve before and after the contamination process. 

\subsection{Period estimation by maximizing Mutual Information}

We propose to use Mutual Information (MI) estimators to detect the underlying period in variable star light curves. In this section we present the rationale behind this proposition. We start by applying the epoch folding transformation for a certain trial period (Eq. \ref{eq-folding}) to the unevenly sampled time instants in order to obtain the phase diagram $\{\phi_i, m_i, \sigma_i\}_{i=1,\ldots,N}$. We assume that the light curve is periodic with an unknown period. The phases $\{\phi_i\}$ correspond to our non-parametric model of the periodicity, while $\{m_i\}$ correspond to our noisy observations. As usual $\{\sigma_i\}$ are the estimated errors on our observations. 

If the light curve is periodic with period $P_T$, then folding with this period will yield the model that best explains our observations. This can be measured by calculating the MI between phases and magnitudes, \emph{i.e.} the amount of information shared by model and observation. We can test several models (foldings) and find the one which maximizes MI to detect the best period. Second-order methods (\emph{e.g.} correlation) are limited to detecting linear relations. More robust periodicity detection methods can be obtained by using MI which overcomes this limitation. Notice that MI requires independent and identically distributed (iid) realizations of the RVs. Although light curves are time series and hence there exist serial correlations in time, these correlations are broken in the phase diagram. Appendix \ref{sec:iid} refers to this issue in detail.

A second interpretation on using MI for periodicity detection lays on MI's definition as the divergence (statistical distance) between the joint PDF and the marginal PDF of the RVs. If the light curve is folded with a wrong period, the structure in the joint PDF will be almost equal to the product of the marginal PDFs, \emph{i.e.} magnitudes are independent of the phases. On the other hand, if the correct period is chosen the joint PDF will present structure that is not captured by the product of the marginals. By maximizing MI we are maximizing the dependency between model and observations. 

Let's denote $M$ and $\Phi$ as the RVs associated to magnitude and phase, respectively. We can estimate the PDF of $M$ given its realizations using KDE as follows
\begin{align}
f_M(m) &= \frac{1}{N} \sum_{i=1}^N \text{G}_{\sqrt{\sigma_i^2+h_m^2}}(m-m_i) \nonumber \\ &= \frac{1}{N} \sum_{i=1}^N \frac{1}{\sqrt{2 \pi (\sigma_i^2 + h_m^2)}} \exp \left( - \frac{1}{2} \frac{(m - m_i )^2}{(\sigma_i^2 + h_m^2)} \right),
\end{align}
where each sample $m_i$ has a bandwidth that incorporates the KDE bandwidth $h_m$ and its given uncertainty $\sigma_i$. As $\Phi$ is a periodic RV we need a periodic kernel to appropriately estimate its PDF. We consider a kernel arising from the Wrapped Cauchy (WC) distribution \citep{jammalamadaka2001topics} and estimate $\Phi$'s PDF as
\begin{align}
f_\Phi(\phi) &= \frac{1}{N} \sum_{i=1}^N \text{WC}_{h_\phi}(\phi-\phi_i) \nonumber \\
&= \frac{1}{2 \pi N} \sum_{i=1}^N \frac{1 - e^{-2 h_\phi}}{1 + e^{-2 h_\phi} - 2 e^{- h_\phi} \cos(2\pi (\phi - \phi_i))},
\end{align}
where $h_\phi \in (0, \infty)$ is the scale of the Cauchy distribution. For $h_\phi \to \infty$ the WC kernel behaves like the circular uniform distribution, while for $h_\phi \to 0$ it concentrates on its mean. The WC kernel is symmetric, translation invariant and closed under convolution\footnote{The convolution of two WC kernels is a WC kernel.} \citep{jammalamadaka2001topics}. Being closed under convolution is desirable because it allow us to compute information potentials efficiently.
The joint PDF of $\Phi$ and $M$ is estimated as
\begin{equation}
f_{\Phi, M}(\phi, m) = \frac{1}{N} \sum_{i=1}^N \text{G}_{\sqrt{\sigma_i^2+h_m^2}}(m-m_i) \cdot \text{WC}_{h_\phi}(\phi-\phi_i),
\end{equation}
because the multiplication of valid kernel functions is also a kernel.

\begin{figure}[t]
\centering
\includegraphics[scale=0.75]{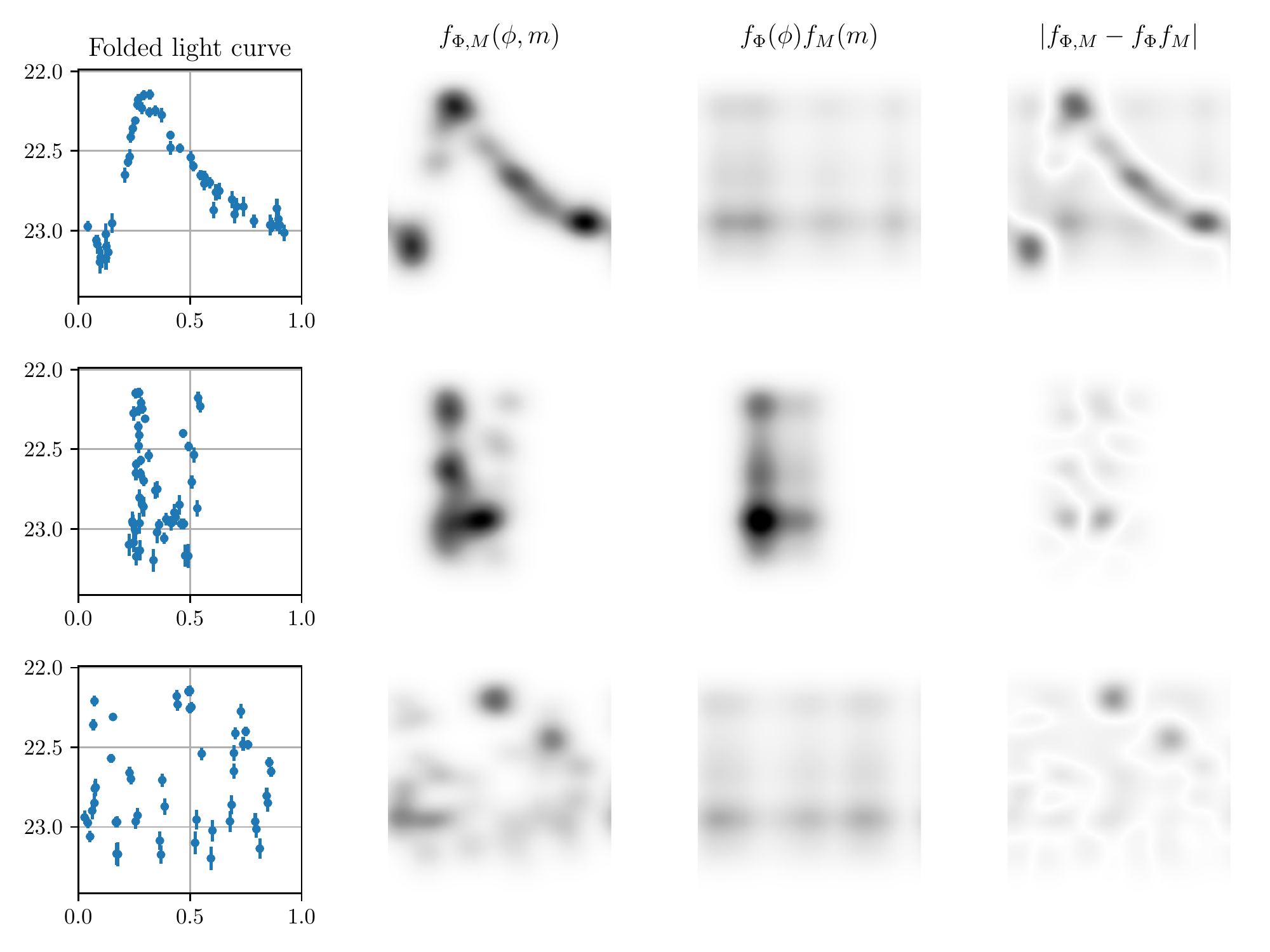}
\caption{\label{fig-examplekde} Folded light curve of a synthetic ab-type RR Lyrae light curve (first column). The light curve is folded using its real period (first row), the sidereal day (second row) and a random period (third row). The second and third columns show the joint PDF and the product of the marginal PDFs of the phases and magnitudes of the light curve. The fourth column is the absolute value of the difference between joint and marginals. A correctly folded light curve produces a large statistical distance between joint and marginals.}
\end{figure}

Fig. \ref{fig-examplekde} shows the estimated joint and product of marginal PDFs of a synthetic ab-type RR Lyrae for three different trial periods, its real period (0.682 days), the sidereal day (0.9973 days) and a random period. By inspecting the PDFs we can see that the difference between the joint (middle column) and marginals (right column) is greater when folding with the true period (first row). 

In Section \ref{sec:review} we reviewed the quadratic MI estimators based on the Euclidean distance (Eq. \ref{eq-QMIED}) and the Cauchy-Schwarz (CS) divergence (Eq. \ref{eq-QMICS}). Our interest on these estimators lays in that they are robust dependency measures and are computed directly from the data bypassing the estimation of PDFs. Computing these estimators requires calculating the information potentials given by Eqs. \ref{eq-vm}, \ref{eq-vj} and \ref{eq-vc}. If we use the Gaussian kernel for the magnitudes and the WC kernel for phases we obtain
\begin{equation} \label{eq-ipm}
    \text{IP}_M = \frac{1}{N^2} \sum_{i=1}^N \sum_{j=1}^{N}  \text{G}_{\sqrt{2h_m^2 + \sigma_i^2+ \sigma_j^2}} \left( m_i-m_j \right),
\end{equation}
\begin{equation} \label{eq-ipp}
    \text{IP}_\Phi = \frac{1}{N^2} \sum_{i=1}^N \sum_{j=1}^{N}  \text{WC}_{2 h_\phi} \left( \phi_i-\phi_j \right),
\end{equation}
\begin{equation} \label{eq-ippm}
    \text{IP}_{\Phi,M} = \frac{1}{N^2} \sum_{i=1}^N \sum_{j=1}^{N}  \text{G}_{\sqrt{2h_m^2 + \sigma_i^2+ \sigma_j^2}} \left( m_i-m_j \right) \text{WC}_{2 h_\phi} \left( \phi_i-\phi_j \right),
\end{equation}
and
\begin{equation} \label{eq-ippxm}
    \text{IP}_{\Phi \times M} = \frac{1}{N} \sum_{i=1}^{N}  \left ( \frac{1}{N}\sum_{j=1}^{N}  \text{G}_{\sqrt{2h_m^2 + \sigma_i^2+ \sigma_j^2}} \left( m_i-m_j \right) \right) \left ( \frac{1}{N} \sum_{j=1}^{N} \text{WC}_{2 h_\phi} \left( \phi_i-\phi_j \right) \right),
\end{equation}
where the Gaussian kernel is used for the magnitudes and the Wrapped Cauchy kernel is used for the phases. Through these potentials we restate the QMI estimators as
\begin{align} \label{eq-QMIEDlc}
\text{QMI}_{ED}(\Phi, M) =   \text{IP}_{\Phi,M} - 2 \text{IP}_{\Phi \times M} + \text{IP}_\Phi \text{IP}_M, 
\end{align} 
and
\begin{align} \label{eq-QMICSlc}
\text{QMI}_{CS}(\Phi, M) =   \log \text{IP}_{\Phi,M} - 2 \log \text{IP}_{\Phi \times M} + \log \text{IP}_\Phi  + \log \text{IP}_M,
\end{align} 
respectively. 

\begin{figure}[t]
\centering
\includegraphics[scale=0.75]{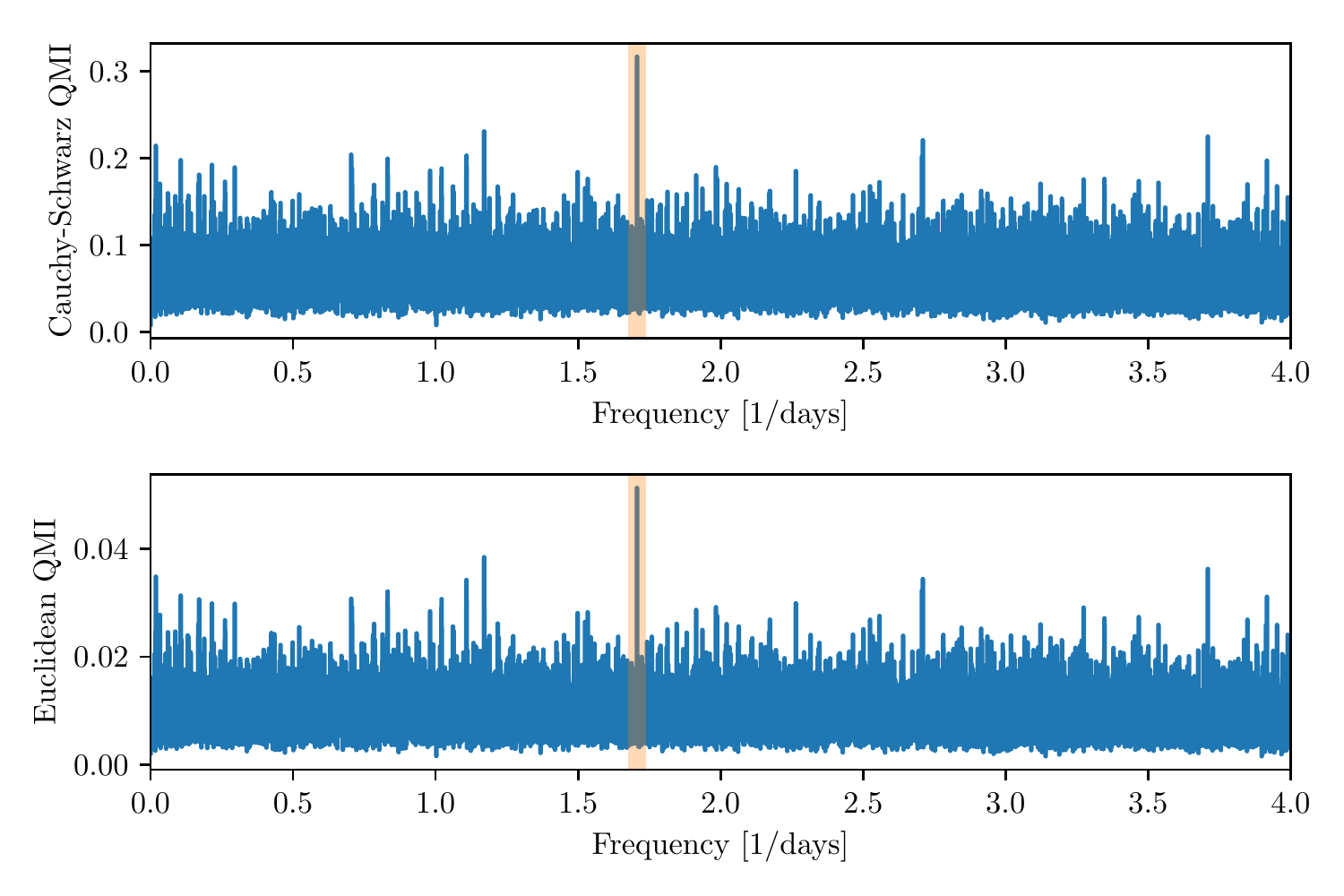}
\caption{\label{fig-exampleqmis} Cauchy-Schwarz and Euclidean QMI estimators as a function of the trial frequency. The underlying period of the light curve is shaded and it corresponds to the highest peak in both cases.}
\end{figure}

The period of a light curve is estimated by maximizing the QMI for a range of trial periods. This yields a QMI periodogram. As an example, in Fig. \ref{fig-exampleqmis} we compute the $\text{QMI}_{CS}$ and  $\text{QMI}_{ED}$ and plot them as a function of frequency for the same light curve used to obtain Fig. \ref{fig-examplekde}. In both cases the underlying period corresponds to the global maximum. In the following section we discuss how to apply the QMI in multi-band light curves.


\subsection{Period estimation in multiband LSST light curves using Quadratic MI}

LSST light curves are characterized for being randomly and sparsely sampled. Methods for period detection that do not aggregate data from all the available bands are likely to fail, specially when few samples per band are available. In this section we show that the QMI periodogram can be easily extended to the case of multi-band light curves. An efficient way to take advantage of the multiple bands is simply to combine the QMI obtained for every single band. But, an average QMI periodogram requires the individual periodograms to be in the same scale. As explained in \citep{principe2010information} the QMI lacks a consistent absolute interpretation because it depends on its parameters, the kernel bandwidths. From \citep{principe2010information} we extract the following conditions regarding comparison between QMI values: (a) the kernel bandwidth has to be selected proportional to the dynamic range of the data and (b) the kernel bandwidth has to be a function of the number of samples and the QMI has to be normalized by its upper bound. The upper bound of the QMI estimators will be studied in the future. In the following experiments we use the same number of samples per band, hence the upper bounds can be ignored.

In our case we have two parameters $h_\phi$ and $h_m$. The former is associated to the phases which are always constrained to $[0, 2\pi]$, \emph{i.e.} the dynamic range of this variable is fixed. QMI is not too sensitive to $h_\phi$ as long as it is not extremely small or large. We have found empirically that $h_\phi = 1$ is a good choice and we keep it constant to make comparisons between QMI values easier. The second bandwidth $h_m$ is more difficult to set as the dynamic range of the magnitudes is not known \emph{a priori}. We consider the plug-in rule from \cite{silverman1986density},
\begin{equation} \label{eq-silver}
h_m = 0.9 \cdot \text{min} ( \sqrt{\text{VAR}[m]}, ~\text{IQR}[m]/1.349) \cdot N^{-1/5},
\end{equation}
where $\text{VAR}[m]$ is the variance of the magnitudes, $\text{IQR}[m]$ is the interquartile range of the magnitudes and $N$ is the number of samples. To avoid overestimation of $h_m$ we use the weighted versions of variance and IQR, with weights $w_i = \sigma_i^{-2}$, $i = 1,\ldots,N$. 
Eq. \ref{eq-silver} complies with the conditions mentioned before. We also explored the Sheather-Jones recurrent estimator \citep{sheather1991reliable} and the more recently proposed diffusion estimator \citep{botev2010kernel} but their performance was not better than Eq. \ref{eq-silver} and their computational cost is higher. In the future local plug-in estimators will be tested. 

\begin{figure}[ht]
\centering
\includegraphics[scale=0.75]{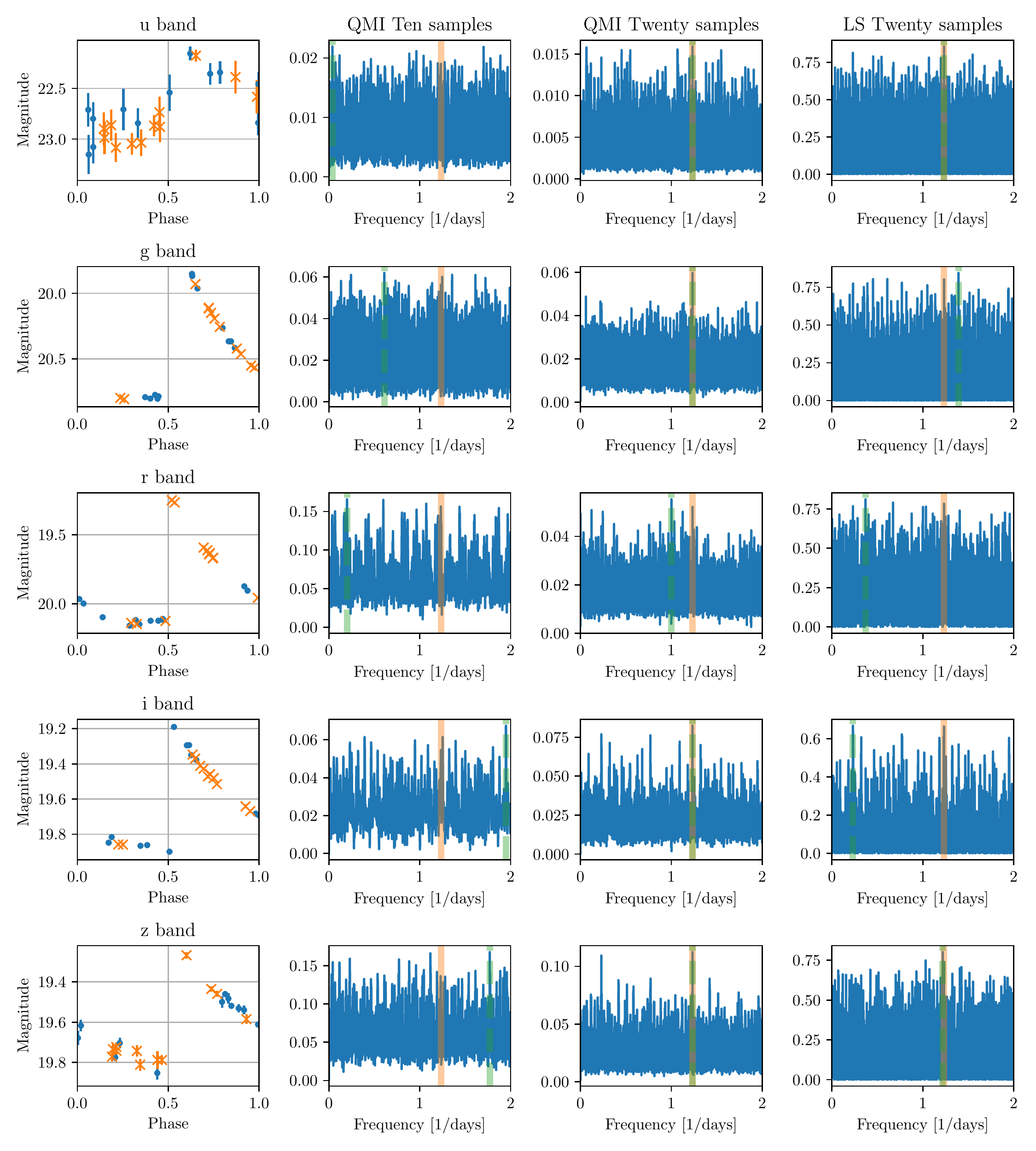}
\caption{\label{fig-exampleqmi-bands} Rows corresponds to different bands (ugriz) of the same synthetic LSST RR Lyrae light curve. The first column shows the light curve folded with its true period, where blue dots and orange crosses correspond to the first ten and second ten samples per band, respectively. Second and third columns correspond to the $\text{QMI}_{ED}$ using ten samples (dots), and twenty samples (dots and crosses) respectively. For reference we include the Lomb-Scargle periodogram (twenty samples) in the fourth column. The true period is shaded with a solid orange line while the maximum of the periodogram is shaded with a dashed green line. }
\end{figure}

\begin{figure}[ht]
\centering
\includegraphics[scale=0.75]{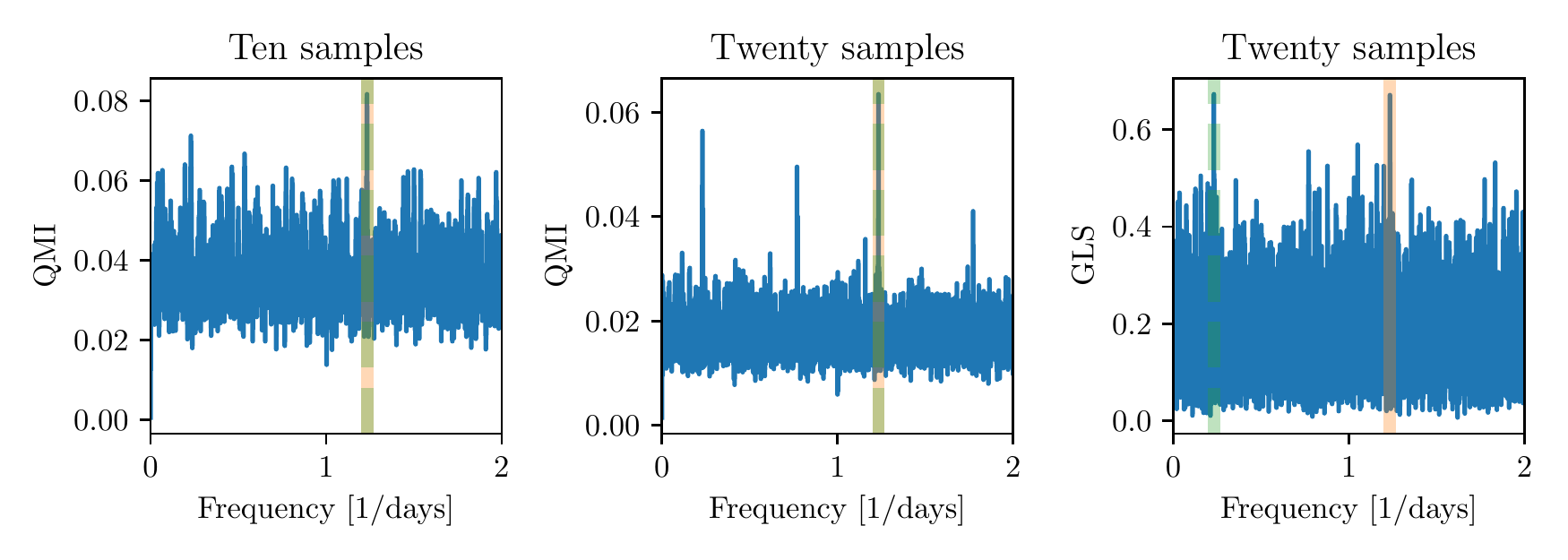}
\caption{\label{fig-exampleqmi-bands-sum} Average QMI across five bands for the example shown in Fig. \ref{fig-exampleqmi-bands}. In both cases the true period is correctly detected. For reference we include the multiband LS periodogram for the 20-sample case.}
\end{figure}

If Eq. \ref{eq-silver} is used for every single band, then the QMI periodograms can be averaged. Fig. \ref{fig-exampleqmi-bands} shows the QMI periodograms for the ugriz bands of a synthetic RR Lyrae light curve. The first column corresponds to the folded light curve. The second and third columns are the Euclidean QMI periodograms using the first ten points and the first twenty points, respectively. Note how using ten points the true period cannot be found in any of the bands. When using twenty points the period is found in the $g$ and $r$ bands only. For reference we show Lomb-Scargle (LS) periodogram for the case of twenty points in the fourth column. 
Fig. \ref{fig-exampleqmi-bands-sum} shows the averaged Euclidean QMI for ten and twenty samples per band. In both cases the true period corresponds to the global maximum of the periodogram. This shows that even if the period is not the maximum in the single band periodograms it can still be found in the combination. The explanation for this is that the true period is likely to appear in all bands although not necessarily as a high peak. Spurious periods due to sampling will not be shared between bands and are de-emphasized in the average periodogram. In the next section we show through extensive experiments that a large gain in performance can be obtained by combining the QMI periodograms.

\section{Results} \label{sec:results}

In this section we test the proposed method using synthetic multi-band LSST light curves. We generate variable star light curves of type ab RR Lyrae (RRab), type c RR Lyrae (RRc) and Cepheids (CEPH) following the procedure described in Section \ref{sec-lsst}. We consider five bands (ugriz) in order to match the original bands of the variability templates. Ten noisy realizations are obtained per generated light curve. This yields a total of 10,000 light curves per variability type.

The QMI estimators are compared to the multiband generalizations of the Lomb-Scargle (LS) and Analysis of Variance (AoV) periodograms. The multiband QMI and AoV methods are implemented in Cython\footnote{\url{http://cython.org/}} and distributed as a python package called \texttt{P4J}\footnote{Available at \url{github.com/phuijse/P4J} and through PyPI.}. For the multiband LS we use the \texttt{gatspy}\footnote{\url{http://www.astroml.org/gatspy/}} python package. All periodograms are run from $0.0$ to $4.0 ~ [1/days]$ with a step size  of $10^{-4} ~ [1/days]$. The kernel bandwidth $h_m$ is set using Eq. \ref{eq-silver} and $h_\phi=1$ in all the experiments. The AOV and generalized LS (GLS) implementations allow for multiharmonic models. We present results using 3 harmonics\footnote{Truncated Fourier series model with fundamental frequency $f_0$, 2 times $f_0$ and 3 times $f_0$ terms.} as this configuration obtains a higher hit rate. For the multiharmonic GLS a conservative regularization term was considered to avoid singularities. All routines are single-core and parallelization is done at time series level. Details on how to set the periodograms using \texttt{P4J} are given in Appendix \ref{sec:python}. 

We consider the period associated to the global maximum of the periodogram as the detected period $P_{D}$. The ability to recover the true period $P_{T}$ is measured in terms of hit rate (HR). We follow \citep{oluseyi2012simulated} and define HR as the number of cases where 
\[
    e_{rel} = \frac{|P_{D} - P_{T}|}{P_{T}^2} < \text{tol},
\] 
divided by the total number of light curves. Tolerance (tol) decreases as a function of $P_T$. Detecting a harmonic or an alias of the true period is considered as a failure. In all experiments the tolerance is set to $1e-3$.

\begin{figure}[t]
\centering
\includegraphics[scale=0.75]{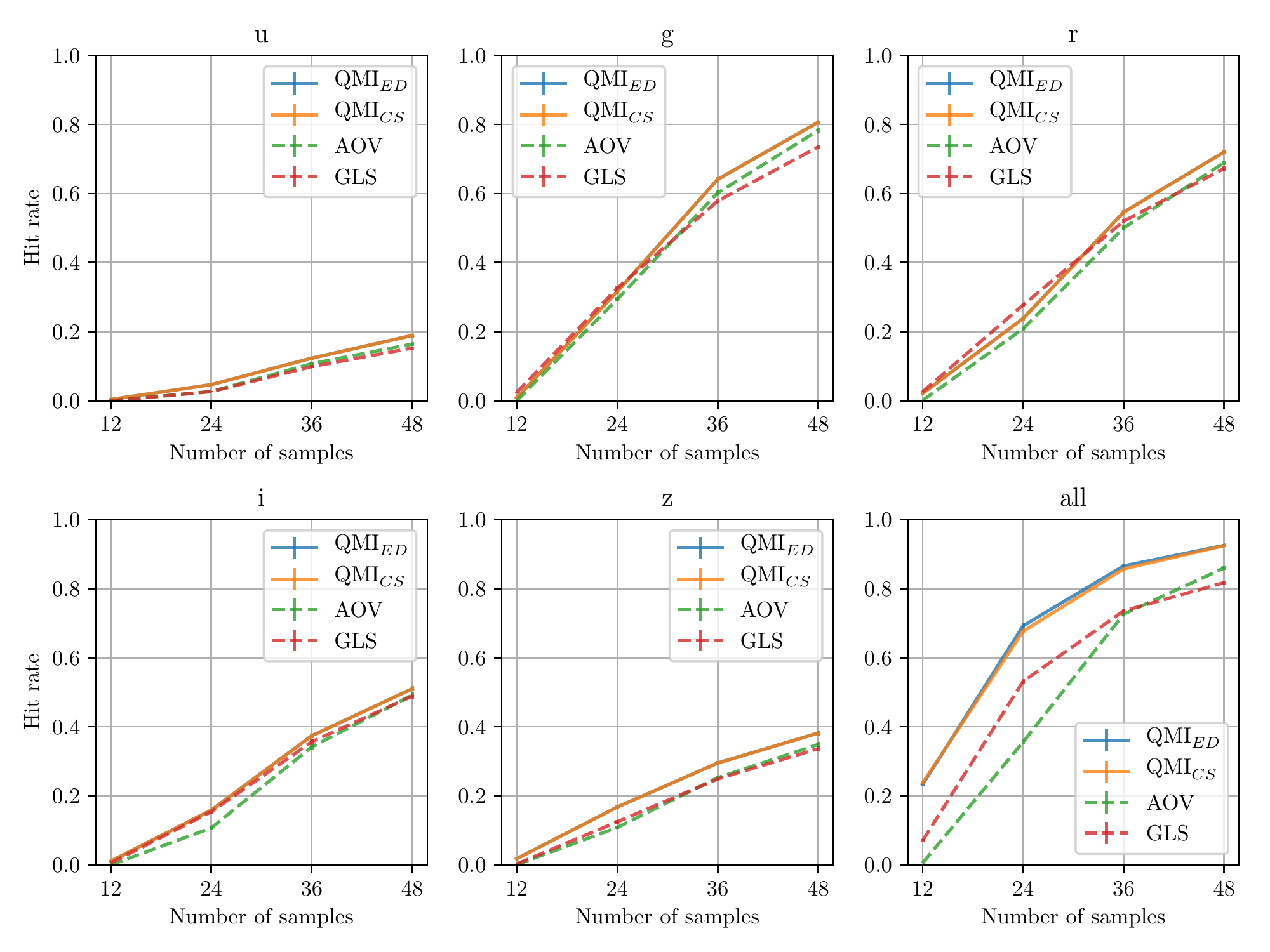}
\caption{\label{fig-res-hr-rrab} Average hit rate for different period detection methods on each band (ugriz) as a function of the number of samples per band. The lower right plot corresponds to the periodograms using all the bands. Each dot is an average of 10,000 synthetic ab-type RR Lyrae light curves. }
\end{figure}

In addition to the multiband periodograms we also evaluate the results on each of the five ugriz bands independently. The robustness against the length of the light curves, \emph{i.e.} the amount of samples required to detect the period, is also studied. Each light curve is evaluated using its first 12, 24, 36 and 48 samples, respectively. Fig. \ref{fig-res-hr-rrab} shows the results of this experiment in the case of RRab templates. Each plot corresponds to one of the ugriz bands while the lower right plot corresponds to the multiband result. In the single-band tests all methods yield a similar performance, but in the multiband test the 2 information theoretic estimators outperform second-order methods. In all tests the difference between the Euclidean and Cauchy-Schwarz QMI hit rates is less than 1\% (their difference is barely noticed in the plots). All methods benefit when aggregating data from the five bands with respect to the best single-band result. Information theoretic (IT) methods yield the largest absolute increase in hit rate when aggregating the data (up to 40\%). Single band best results are obtained in \emph{g} which is expected as RR Lyrae are inherently more variable in this filter. 

\begin{figure}[t]
\centering
\includegraphics[scale=0.75]{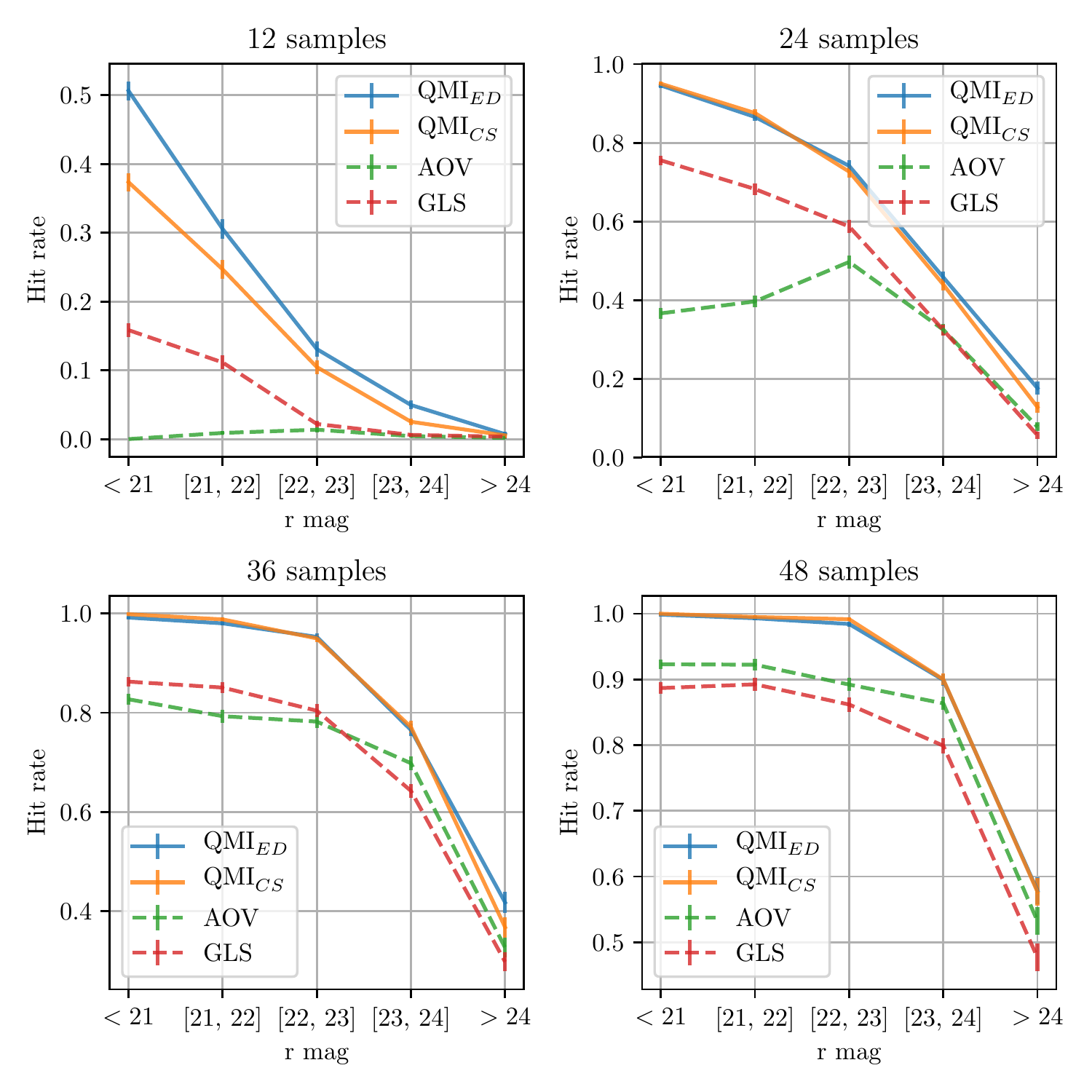}
\caption{\label{fig-res-hr-rrab-mag} Average hit rate for different period detection methods as a function of the r-band magnitude. Each plot corresponds to a different light curve length. All light curves correspond to ab-type RR Lyrae. LSST will produce an average of 18.4 samples per year in the r-band.} 
\end{figure}

Fig. \ref{fig-res-hr-rrab-mag} shows the multiband hit rates averaged over different ranges of the magnitude in the \emph{r}-band. Each plot corresponds to a different light curve length (sample size). Signal-to-noise ratio (SNR) decreases with magnitude. As expected HR increases with light curve length and decreases with magnitude for all methods. Both QMI estimators have a similar performance, except in the 12 samples case where the Euclidean QMI performs better than CS QMI suggesting that the former might be more robust to low sample size. QMI estimators outperform second-order methods in all cases. This is more noticeable for shorter light curves, with the absolute HR margin growing from 10\% to 30\% (Euclidean QMI vs multiband GLS). This shows that QMI estimators can detect the true period faster (in survey time) than second-order methods. The multiband AoV performs slightly better than the multiband GLS in the 48 samples case. On the other hand GLS performs considerably better than AoV at shorter light curve lengths and brighter magnitudes. In the 24-sample case the performance of AoV decreases with SNR, AoV tends to recover harmonics of the true period more frequently in this regime. In all cases, the difference in hit rate between methods decreases when approaching the \emph{r}-band $5\sigma$ limit of 24.5. 

\begin{figure}[t]
\centering
\includegraphics[scale=0.75]{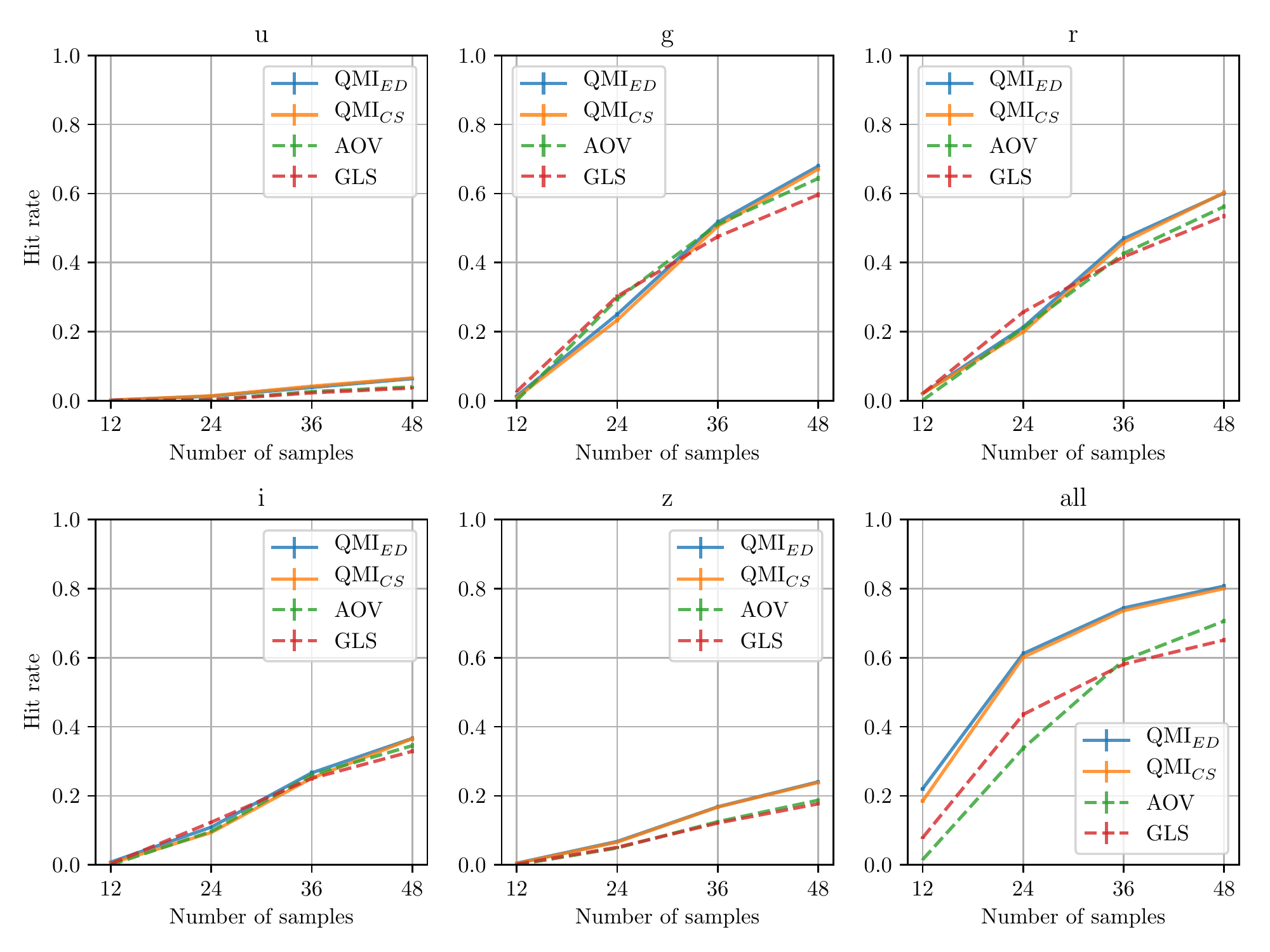}
\caption{\label{fig-res-hr-rrc} Average hit rate for different period detection methods on each band (ugriz) as a function of the number of samples per band. The lower right plot corresponds to the periodograms using all the bands. Each dot is an average of 10,000 synthetic c-type RR Lyrae light curves. }
\end{figure}

\begin{figure}[t]
\centering
\includegraphics[scale=0.75]{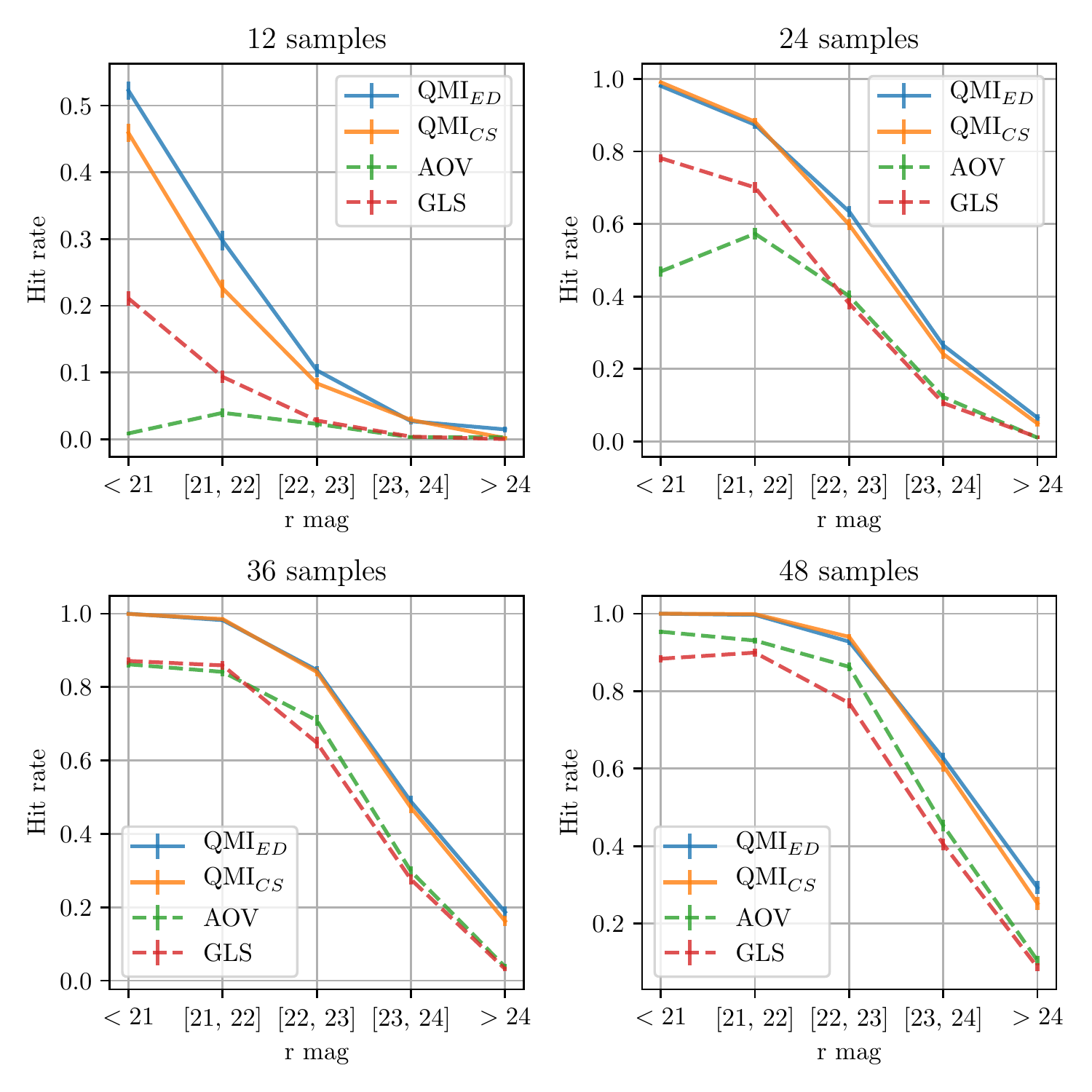}
\caption{\label{fig-res-hr-rrc-mag} Average hit rate for different period detection methods as a function of the r-band magnitude. Each plot corresponds to a different light curve length. All light curves correspond to c-type RR Lyrae.}
\end{figure}

Fig. \ref{fig-res-hr-rrc} shows the results obtained using the RRc templates. Again we can see that single-band results are only marginally different between methods. As with RRab, single-band best results are obtained in the \emph{g}-band. The lower right plot shows the multiband results. In the multiband case the QMI based methods see an increase in hit rate between 15\% and 40\% with respect to single-band best results, outperforming second-order methods. Fig. \ref{fig-res-hr-rrc-mag} shows the multiband results in more detail. The QMI methods perform better than second-order methods in all cases, and the difference in HR grows for shorter light curves. Both QMI estimators perform similarly except in the 12 sample case where the Euclidean estimator performs better.

\begin{figure}[t]
\centering
\includegraphics[scale=0.75]{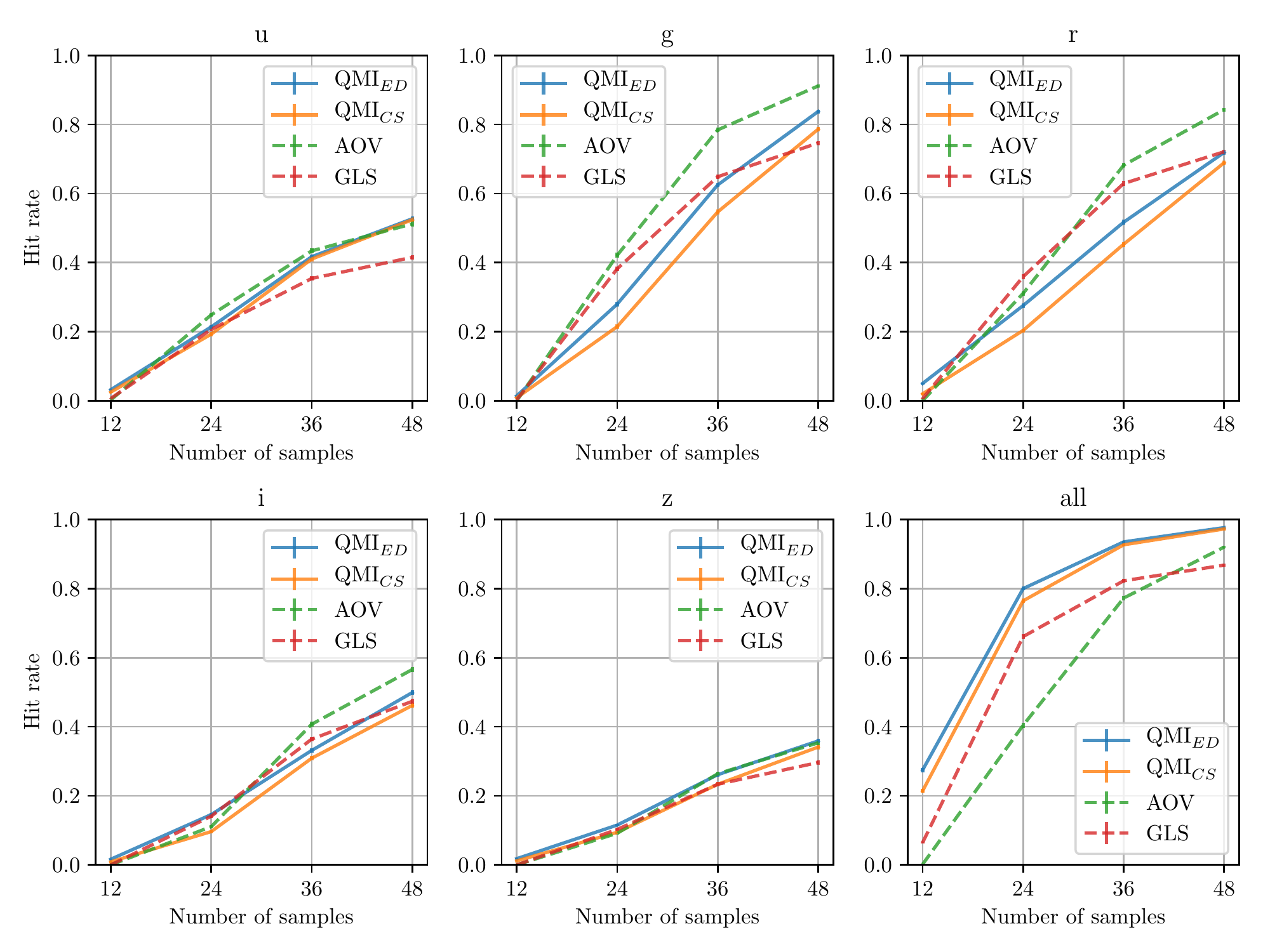}
\caption{\label{fig-res-hr-ceph} Average hit rate for different period detection methods on each band (ugriz) as a function of the number of samples per band. The lower right plot corresponds to the periodograms using all the bands. Each dot is an average of 10,000 synthetic Cepheid light curves. }
\end{figure}

\begin{figure}[t]
\centering
\includegraphics[scale=0.75]{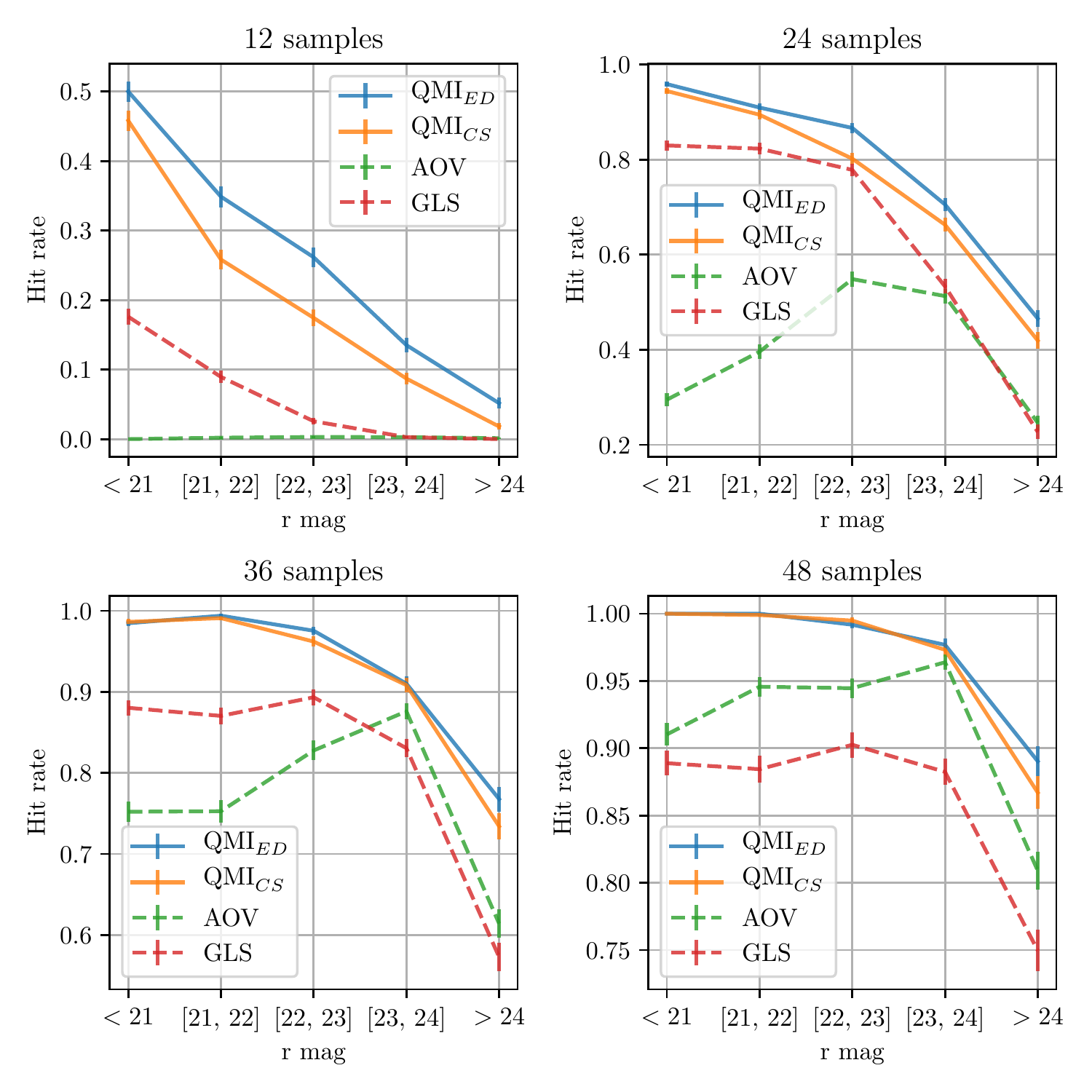}
\caption{\label{fig-res-hr-ceph-mag} Average hit rate for different period detection methods as a function of the r-band magnitude. Each plot corresponds to a different light curve length. All light curves correspond to Cepheids.}
\end{figure}

Fig. \ref{fig-res-hr-ceph} shows the results obtained with the CEPH templates. QMI estimators outperform their competitors in the multiband case, but perform worse in the single \emph{g} and \emph{r} band cases. Interestingly, there is little gain when aggregating bands for the AoV periodogram. On the other hand, QMI methods see an absolute increase in HR from 10\% to 50\%. The Euclidean QMI performs slightly better than the CS QMI in all tests. Fig. \ref{fig-res-hr-ceph-mag} shows the multiband results in greater detail. The Euclidean QMI performs better than the CS QMI when sample size decreases. QMI methods perform better than second-order methods, and again this is more evident when sample size decreases (shorter light curves). The GLS perform better than AOV except in the 48 samples case. Once again, we note a strong tendency of AoV to recover a harmonic of the true period for brighter magnitudes and smaller sample sizes. 

Table \ref{tab-times} shows the computational time required to calculate a complete periodogram using our library on time series of different lengths (time is an average of 100 repetitions). Computational time is measured on a Intel i5-4460 CPU at 3.20GHz. Computational time is on the same order of magnitude, but due to the increased computational complexity of QMI estimators they scale worse with number of samples. We are working on approximations of the information potential estimator to reduce the computational time in the case of dense light curves. 

\begin{table}[t]
\centering
\begin{tabular}{ c c c }
 Number of samples & Euclidean QMI & AOV  \\ \hline
 12 & $0.074 \pm 0.012$ & $0.078 \pm 0.011$  \\  
 24 & $0.185 \pm 0.013$ & $0.123 \pm 0.017$  \\
 36 & $0.298 \pm 0.031$ & $0.167 \pm 0.018$  \\
 48 & $0.492 \pm 0.028$ & $0.225 \pm 0.025$  
\end{tabular}
\caption{\label{tab-times} Computational time (seconds) using \texttt{P4J} to compute a periodogram on a single time series (averaged over 100 repetitions). AOV is computed with 3 harmonics.}
\end{table}



\section{Conclusion and Future work} \label{sec:conclusion}

In this paper we have proposed an estimator of the quadratic mutual information (QMI) for estimating periods in light curves. By maximizing the QMI between the phases and the magnitudes of a light curve the underlying period can be estimated. Contrary to second-order methods, the QMI extracts information from the whole PDF, is not restricted to linear relations between variables and is more robust to non-Gaussian (heavy-tailed) noise. Efficient Cython implementations of the methods presented in this paper are freely available through \texttt{github} and \texttt{PyPI}.

We have applied the QMI for period estimation in sparse multi-band light curves of variable stars generated with the LSST simulation tools. The OpSim and CatSim tools allow us to build a database of realistic synthetic light curves with multi-band pointings, cadence and noise distribution as expected by the LSST. Our results show that the QMI outperforms the multiband generalizations of the Lomb-Scargle and Analysis of Variance periodograms for all variability types. The QMI is efficient at aggregating data from several sparsely-sampled bands, presenting an absolute hit rate increase up to 50\% with respect to best single-band results. We have observed that the performance gap with second-order methods is more noticeable when the length of the light curve decreases (smaller sample sizes). The multiband QMI is more robust to noise and it can detect the true period faster (survey time) than second-order methods. 

In our proposition we combine the single-band QMI periodograms allowing us to detect the period even when individual periodograms cannot. This however does not exploit the interaction between bands directly. We recognize an extension to this proposition that involves calculating QMI cross-products between bands. Although we note that this would increase the computational complexity, we have preliminary results that show that using these cross-products allows for even more robustness against low sample size and noise.

One weakness of the proposed estimator is that it scales quadratically with the number of samples, making it expensive to compute for dense data (more than 100 samples per band). We plan to include a Fast-Gauss transform implementation of the information potential estimator in our library in the near future to partially solve this. We will also study new ways to estimate MI that might be more efficient such as \cite{giraldo2015measures} propositions.

Future work also includes a more profound analysis of the differences between the Euclidean and Cauchy-Schwarz QMI, and studying the upper bounds of these estimators. We expect to develop relative QMI estimators that allow us to compare results between different light curves, which is key to develop statistical criteria based on the QMI distribution to avoid the cost of case-by-case bootstrap analysis. In this work we focused on quadratic (order 2) entropy and MI estimators. In the future we will test MI estimators of different orders and study their properties.

\section*{Acknowledgement}
Pablo Huijse (P.H.) acknowledges support from FONDECYT through grant \textnumero 1170305. Pablo A. Est\'evez (P.E.) acknowledges support from FONDECYT through grant \textnumero 1171678. Francisco F\"orster (F.F.) acknowledges support from FONDECYT through grant \textnumero 3110042 and from Basal Project PFB-03. P.H., P.E. and F.F. acknowledge support from CONICYT through the Programme of International Cooperation project DPI20140090 and from the Chilean Ministry of Economy, Development, and Tourism's Millennium Science Initiative through grant IC12009, awarded to The Millennium Institute of Astrophysics, MAS. Andrew J. Connolly acknowledges partial support by the U.S. Department of Energy, Office of Science, under Award Number DE-SC-0011635, from the DIRAC Institute and the LSST, and from the NSF through awards AST-1409547 and AST-1715122. Powered@NLHPC: This research was partially supported by the supercomputing infrastructure of the NLHPC (ECM-02). Part of this work was done under the Harvard-Chile data science school.

\appendix

\section{Information Theoretic Learning} \label{sec:ITL}

Information Theoretic Learning (ITL) \citep{principe2000, principe2010information} is a framework to bring information theoretic criteria into machine learning (ML) methods. Traditionally, ML methods are trained via optimizing a second-order loss function, \emph{e.g.} the mean square error (MSE) and correlation. In ITL these quantities are replaced by information theoretic criteria that describe the probability density function (PDF), \emph{e.g.} entropy and mutual information. ITL criteria extract more information from data improving the performance of the methods. By going beyond the second-order moment ITL criteria gain robustness in realistic scenarios where the Gaussianity assumption does not hold, \emph{e.g.} under the presence of heavy-tailed noise and outliers. 

In ITL a strong emphasis is given to the estimation of these quantities directly from data in a non-parametric way. As an example consider the ITL estimation of Renyi's second order generalization $H_{2} (X)$ of Shannon's entropy  \citep{principe2010information}  of a continuous RV defined as 
\begin{equation} \label{eq-r2-def}
    H_{2} (X) = - \log \int f_X(x)^2 \,dx,
\end{equation}
where $f_X(x)$ is the RV's PDF. Assuming that we have $\{x_i\}_{i=1,\dots,N}$ realizations of the RV its PDF can be computed using a kernel density estimator (KDE)
\begin{equation} \label{eq-kde}
    f_X(x) = \frac{1}{N} \sum_{i=1}^N \text{G}_h \left( x-x_i\right) = \frac{1}{N\sqrt{2\pi}h} \sum_{i=1}^N \exp \left( \frac{\|x-x_i\|^2}{2h^2} \right),
\end{equation}
where $\text{G}_h(\cdot)$ is the Gaussian kernel with bandwidth $h$. By replacing Eq. \ref{eq-kde} in Eq. \ref{eq-r2-def} and then using the Gaussian convolution property\footnote{The convolution of two Gaussian functions is also a Gaussian.} we obtain
\begin{align} \label{eq-r2-kde}
    H_{2} (X) &= - \log \frac{1}{N^2} \int  \sum_{i=1}^N \sum_{j=1}^N  \text{G}_h \left( x-x_i \right)  \text{G}_h \left(x-x_j \right) \,dx, \nonumber \\
    &= - \log \frac{1}{N^2}  \sum_{i=1}^N \sum_{j=1}^N  \text{G}_{\sqrt{2}h} \left( x_i-x_j \right) = - \log \text{IP}_X,
\end{align}
where $\text{IP}_X$ is the Information Potential (IP), an estimator of the expected value of the PDF of $X$ \citep{principe2010information}, although it is estimated directly from the data samples bypassing the estimation of the PDF. Other symmetric and translation-invariant kernels can be used, but it is convenient to use kernels which are closed under convolution. 

\section{Testing the iid hypothesis in the phase diagram} \label{sec:iid}

\begin{figure}[t]
\centering
\includegraphics[scale=0.75]{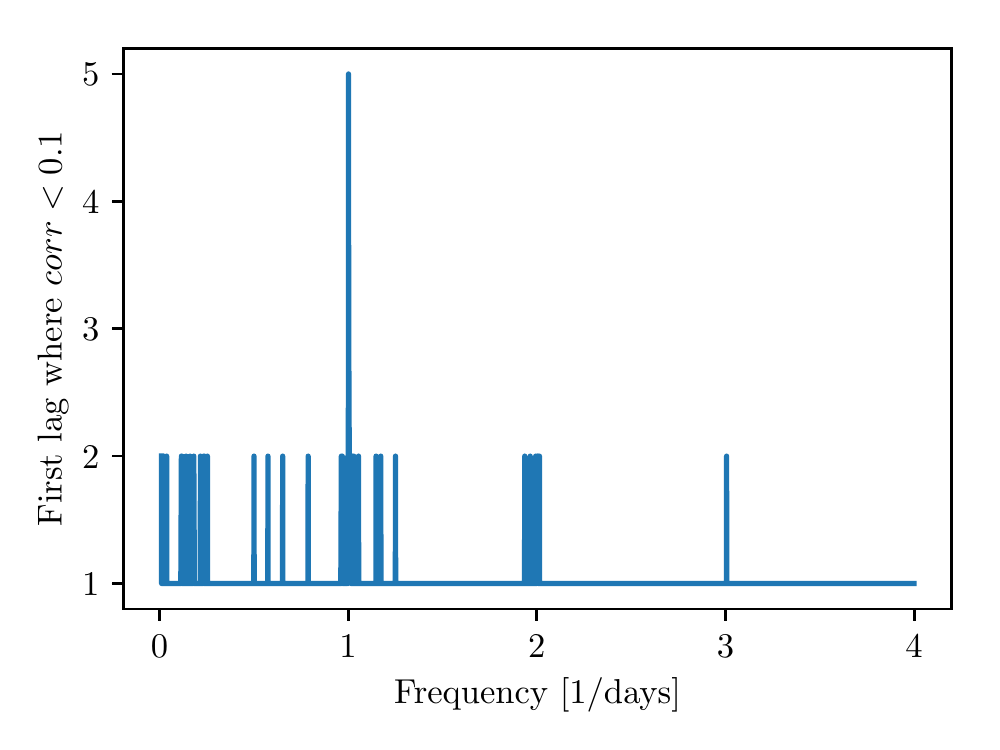}
 \includegraphics[scale=0.75]{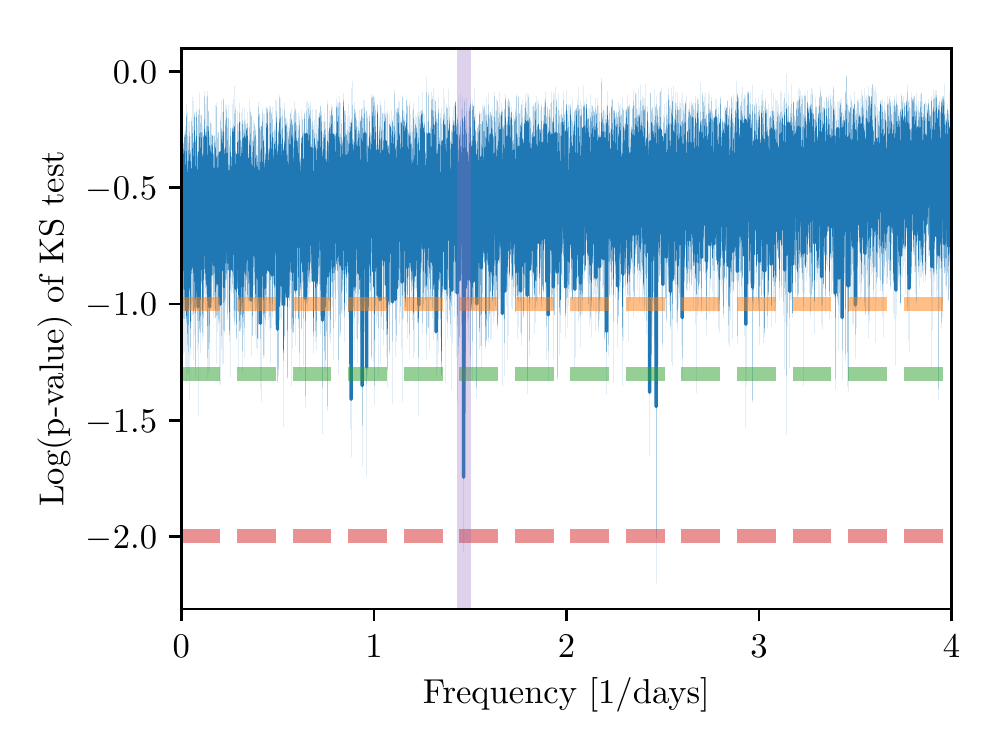}      
\caption{\label{fig-iid} (Left) First lag for which the slotted autocorrelation is less than 0.1 as a function of frequencies (foldings). In most cases autocorrelation has reached this value at the first lag. Harmonics of the sidereal day are a notable exception of this. (Right) Average logarithm of the p-value of a KS test across data partitions for different frequencies (foldings). The shaded area correspond to the errorbars. The dotted line mark the 10\%, 5\% and 1\% significance. In most cases the null hypothesis cannot be rejected with a significance smaller than 10\%. A notable exception is the period of the light curve (shaded), which rejects the null with a significance smaller than 5\%.}
\end{figure}

For the estimation of mutual information (MI) we assume that the samples from our joint PDF $f(\Phi, M)$ are independent and identically distributed, \emph{i.e.} all realizations come from the same continuous distribution and no serial correlations exists between realizations. Light curves are time series so we expect to find temporal correlations, although sampling is pseudo-random and does not obey Nyquist's theorem \citep{eyer1999variable}. Phase is a function of time and period, and several periods are tested per light curve. If the period is not related to the underlying periodicity of the data the phase diagram is filled uniformly and serial correlations in the joint space are broken. This is shown in Fig. \ref{fig-iid} for a periodic light curve. The plot on the left shows that for most frequencies (foldings) the slotted autocorrelation drops very fast. Assuming that the light curve is stationary (no trends) we can partition the phase-magnitude space in equally sized bins and compute a two-dimensional Kolmogorov-Smirnov (KS) test \citep{fasano1987multidimensional}\footnote{Implemented at \url{github.com/syrte/ndtest}.} to consider the null hypothesis that the binned distributions are equal. Fig. \ref{fig-iid} (right) shows the logarithm of the average p-value as a function of frequency. In the majority of cases we cannot reject the null at 10\% significance. We evaluate a subset of 1,000 light curves with different periods and found these results to be consistent. This explains why MI works well in practice when applied to the folded data.

\section{Using the P4J python library} \label{sec:python}

Listing \ref{lst:P4J} demonstrates how to compute the QMI periodogram using the python P4J library (\url{github.com/phuijse/P4J}). The data used for this paper and multiband evaluation scripts based on P4J can be found at \url{github.com/phuijse/LSST_simulations}. 

\begin{lstlisting}[language=Python, label={lst:P4J}, caption="P4J demonstration"] 
import P4J
# Assuming that mjd, mag and err are N-length numpy arrays
# Using the Euclidean QMI periodogram
my_per = P4J.periodogram(method="QMIEU", debug=False)
# By default silverman's rule is used and hp=1
my_per.set_data(mjd, mag, err, whitten=False)
my_per.frequency_grid_evaluation(fmin=0.0, fmax=4.0, fresolution=1e-4)
# You may want to finetune the estimations around the maxima
my_per.finetune_best_frequencies(fresolution=1e-5, n_local_optima=10)
# If you want the whole periodogram
freq, per = my_per.get_periodogram()
# If you only want  to retrieve the best frequencies
fbest, pbest = my_per.get_best_frequencies()
# Other available methods are QMICS, MHAOV, PDM1 and LKSL

\end{lstlisting}

\bibliography{references}

\end{document}